\documentclass[%
 reprint,
 amsmath,amssymb,
 aip,
 jcp,
]{revtex4-2}

\usepackage{graphicx}
\usepackage{dcolumn}
\usepackage{bm}
\usepackage[dvipsnames]{xcolor}

\begin{document}

\title{Nucleation rates from small scale atomistic simulations and transition state theory}

\author{Kristof M. Bal}
  \email{kristof.bal@uantwerpen.be}
  \affiliation{Department of Chemistry and NANOlab Center of Excellence, University of Antwerp, Universiteitsplein 1, 2610 Antwerp, Belgium}

\date{\today}

\begin{abstract}
The evaluation of nucleation rates from molecular dynamics trajectories is hampered by the slow nucleation time scale and impact of finite size effects.
Here, we show that accurate nucleation rates can be obtained in a very general fashion relying only on the free energy barrier, transition state theory (TST), and a simple dynamical correction for diffusive recrossing.
In this setup, the time scale problem is overcome by using enhanced sampling methods, in casu metadynamics, whereas the impact of finite size effects can be naturally circumvented by reconstructing the free energy surface from an appropriate ensemble.
Approximations from classical nucleation theory are avoided.
We demonstrate the accuracy of the approach by calculating macroscopic rates of droplet nucleation from argon vapor, spanning sixteen orders of magnitude and in excellent agreement with literature results, all from simulations of very small (512 atom) systems.
\end{abstract}


\maketitle

\section{Introduction}

First order phase transitions are initiated by a nucleation event, in which a small embryo of a thermodynamically favored phase is formed within a bulk metastable phase.
Nucleation is an inherently difficult process to study.
In principle, the nanoscale dimensions of the critical nucleus make molecular dynamics (MD) simulations a natural choice to probe the nucleation process.
The rare event nature of critical nucleus formation, which may take seconds or longer, however puts it well beyond the MD time scale.

The difficulties associated with nucleation simulations are nicely illustrated by one of the simplest nucleation processes, namely, the formation of a liquid droplet in argon vapor.
Even here, direct MD simulations can only capture nucleation events at very high supersaturations~\cite{Chkonia2009} or in very expensive massively parallel large-scale calculations.~\cite{Diemand2013}
In addition, the use of computationally efficient small simulation cells introduces significant finite size artifacts.~\cite{Wedekind2006}
Moreover, indirect rate calculations based on classical nucleation theory (CNT) may be in error by several orders of magnitude.~\cite{Diemand2013} 

Recently, accelerated molecular dynamics approaches have started to address the key issues in this field.~\cite{Salvalaglio2016,Tsai2019}
Slow argon droplet nucleation events can be observed in direct MD simulations when an external bias potential is applied.
Under certain conditions, it is then possible to quantitatively correct for the impact of the bias potential on the apparent (shortened) nucleation time.~\cite{Voter1997,Tiwary2013}
This way, trajectories corresponding to physical nucleation times up to $\tau = 10^4$~s have been sampled.
This methodology is in principle highly generic because, besides the simulation model itself, no specific mechanistic assumptions are made.
In addition, by using concepts from CNT, estimates of macroscopic nucleation rates were obtained \textit{a posteriori} by correcting nucleation times from small-scale simulations.~\cite{Salvalaglio2016}
Such accelerated MD approaches have however not yet been applied to other types of nucleation problems.

Yet, the enhanced sampling methods on which recent accelerated MD work has been based have already been broadly applied to different types of phase transitions in atomistic simulations.
Recent examples of such studies include melting,~\cite{Samanta2014} solid--solid transitions,~\cite{Gimondi2017,Rogal2019} crystallization from the liquid,~\cite{Quigley2009,Piaggi2017,Pipolo2017,Zhang2019,Piaggi2020,Karmakar2021} and crystallisation from solution.~\cite{Salvalaglio2015,Karmakar2019,Fukuhara2021}
Such approaches however do not directly produce nucleation rates, and rather aim to reconstruct the free energy surface (FES) of the nucleation process.

The FES concept unifies the description of thermodynamics across systems, avoiding any process-specific theories: an appropriate set of low-dimensional order parameters is the only required system-dependent information.
Calculated FES for nucleation have therefore primarily been used to investigate the thermodynamic aspects of phase transitions.

The FES does, however, in principle also encode kinetic information.
It is possible to obtain free energy barriers and, thus, calculate rates using transition state theory (TST), at least for chemical reactions.~\cite{Bal2020}
Would this also be possible for nucleation, thus unifying rate calculation within a single framework?
Indeed, dedicated ``extraordinary rate theories~\cite{Peters2015}'' for specific processes tend to share many common aspects and are ultimately equivalent.

Even though a large portfolio of tools has already been applied to the calculation of nucleation rates---with recent studies of various processes having employed seeding approaches,~\cite{Espinoza2016,Zimmermann2018} transition path sampling (TPS),~\cite{Arjun2021} transition interface sampling (TIS),~\cite{Arjun2020,Menon2020} and forward flux sampling (FFS)~\cite{Wang2009,Haji-Akbari2015,Sosso2016,Jiang2018}---accurate calculation of realistic nucleation rates is a formidable challenge in general.
Computed nucleation rates in any system are highly sensitive to every methodological aspect or approximation in sometimes non-obvious ways.~\cite{Blow2021}
Moreover, different approaches to the calculation of nucleation rates can sometimes disagree quite strongly.~\cite{Diemand2013,Diemand2014,Cheng2017,Cheng2018}
Any new methodological perspective might therefore be valuable to understand such discrepancies.

In this manuscript, we demonstrate that highly accurate nucleation rates can be calculated within the generic workflow of a free energy calculation.
It is not necessary to rely on mechanistic assumptions~\cite{Schenter1999} or expression derived from CNT,~\cite{Auer2001,Auer2004} nucleation events need not be explicitly sampled from dynamical trajectories, and it is possible to account for finite size effects in a natural manner.
In order to demonstrate the accuracy of our approach, we first validate its rate estimates for droplet nucleation from argon vapor in a small simulation cell over a wide range of supersaturations.
We then show how, using the same small-scale system, this workflow also allows to calculate rates free of finite size errors.

\section{Methodology}

\subsection{TST rates for droplet nucleation}

In order to generate a FES of an arbitrary process, one must first identify at least one suitable collective variable (CV) $\chi = \chi (\mathbf{R})$ that is a function of the system coordinates $\mathbf{R}$ and that can distinguish all states of interest.
This CV will also serve as a candidate reaction coordinate of the process.
Recently, we have demonstrated that, whenever such a FES $F (\chi)$ is available for a process $A \rightarrow B$, its TST rate $k^\mathrm{TST}$ can be calculated unambiguously.~\cite{Bal2020}
For an \emph{arbitrary} choice of the reaction coordinate $\chi$ one can derive that the total flux through a dividing surface $\chi = \chi^*$ in the configuration space $\mathbf{R}$ is equal to
\begin{equation}
  k^\mathrm{TST} = \sqrt{\frac{1}{2 \pi \beta m}} \frac {\displaystyle\int  \mathrm{d} \mathbf{R} \left | \nabla \chi \right |_{\chi = \chi^*} \cdot \delta[ \chi^* - \chi(\mathbf{R})] \, e^{-\beta U (\mathbf{R})}}{\displaystyle\int \mathrm{d} \mathbf{R} \, H[ \chi^* - \chi(\mathbf{R})] \, e^{-\beta U (\mathbf{R})}} \label{eq:kTST},
\end{equation}
in which $U (\mathbf{R})$ is the potential energy of the system.
$|\nabla n|$ is the norm of the gradient of $\chi (\mathbf{R})$ with respect to all coordinates $\mathbf{R}$ and serves as a gauge correction to ensure invariance of the rate with respect to the parametrization of $\chi$.
$\beta = (k_B T)^{-1}$, $k_B$ the Boltzmann constant, $T$ the temperature, and $m$ the mass of the nucleating particles.
The step function $H$ and delta function $\delta$ are used to select configurations belonging to the initial metastable state ($\chi < \chi^*$) and dividing surface ($\chi = \chi^*$) respectively.
The dividing surface $\chi = \chi^*$ can also be referred to as the location of the transition state (TS).
This expression can be recast in terms of the free energy surface $F(\chi)$ (or $G(\chi)$):
\begin{equation}
  k^\mathrm{TST} = \frac{\langle |\nabla \chi| \rangle_{\chi=\chi^*}}{\sqrt{2 \pi \beta m}} e^{-\beta (F(\chi^*) - F_A)} ,
\end{equation}
in which state $F_A$ is the integrated free energy of state $A$, which we have defined as all configurations for which $\chi < \chi^*$:
\begin{equation}
  F_A = -\frac{1}{\beta} \ln \int_{\chi < \chi^*}  \mathrm{d} \chi \, e^{-\beta F(\chi)} . \label{eq:FA}
\end{equation}
We can now rewrite the expression for $k^\mathrm{TST}$ to be equivalent to the well-known Eyring formula:
\begin{equation}
  k^\mathrm{TST} = \frac{1}{h \beta} e^{-\beta \Delta^\ddagger F} , \label{eq:Eyring}
\end{equation}
in which $h$ is the Planck constant.
To do this, we must define the free energy barrier $\Delta^\ddagger F$ as:
\begin{equation}
  \Delta^\ddagger F_{A \rightarrow B} = F(\chi^*) + \frac{1}{\beta} \ln \frac{\langle |\nabla \chi| \rangle^{-1}_{\chi=\chi^*}}{h} \sqrt{\frac{2 \pi m}{\beta}} - F_A. \label{eq:barF} 
\end{equation}
As we have argued before, besides ensuring compatibility with the Eyring expression, this definition of $\Delta^\ddagger F$ also has a physical significance: It measures the probability of generating a configuration $\chi = \chi^*$ from the full ensemble of $A$ states, i.e., states for which $\chi < \chi^*$.~\cite{Bal2020}

Note that we have made no assumptions whatsoever about the mechanism or nature of the $A \rightarrow B$ transition, or put any requirements on $\chi$.
To be useful, $\chi$ of course has to be able to properly discriminate states $A$ and $B$, and parametrize an appropriate dividing surface between the two.

In a condensation process state $A$ is the vapor phase $g$, and $B$ is the liquid $l$.
For droplet nucleation, the number of liquid atoms (the ten Wolde--Frenkel parameter) $n$~\cite{tenWolde1998} has been a common choice.~\cite{Chkonia2009,Salvalaglio2016,Tsai2019}
An atom is considered liquid when it has more than 5 close neighbors.~\cite{tenWolde1998}

Thus, we choose $\chi = n$.
It must be noted that $n$, as defined in this work, does not strictly count the number of atoms inside the largest droplet.
It is rather a measure of the number of highly coordinated atoms.
Low-coordinated atoms also make a (small) contribution to $n$, even if they have no direct neighbors, while atoms at the surface of a droplet may not be fully counted.
This is a consequence of requiring $n$ to be a computationally convenient continuous function that can be used to induce transitions from gas to liquid, and back.
As a result, $n$ is also not a valid definition of cluster size within the context of classical nucleation theory because it does not strictly count the number of atoms inside the largest cluster only.~\cite{Cheng2017}

None of this matters much from a TST perspective.
$n = n (\mathbf{R})$ is ultimately just some order parameter that exfoliates configuration space and parametrizes a dividing surface $n = n^*$.
$n^*$ is in this context defined as the value of $n$ that maximizes the geometric free energy surface $F^G(n)$~\cite{Hartmann2007}:
\begin{equation}
  F^G (n) = F (n) - \frac{1}{\beta} \ln \langle |\nabla n| \rangle_{n(\mathbf{R}) = n} \label{eq:GFES} .
\end{equation}
This is mathematically equivalent to variationally minimizing the TST rate Eq.~\eqref{eq:kTST}, because it is the classical upper bound of the true rate.
$F^G (n)$ can be obtained simultaneously with the standard FES $F(n)$ by reweighting.~\cite{Bal2020}

We sometimes refer to $n^*$ as the ``critical nucleus size'' although it is, strictly speaking, not the size of the actual critical nucleus within macroscopic CNT for the reasons outlined above.
Whether or not we can identify the critical nucleus---or if it even exists---is however irrelevant.
After all, we could just as well have used a more advanced reaction coordinate that also accounts for cluster shape~\cite{Tsai2019} or one that has an even less pronounced connection to the nucleus size, such as a generic measure of global order.~\cite{Piaggi2017}
Any possibility to interpret the reaction coordinate in terms of cluster size is then lost, but we can still proceed to calculate the TST rate as described previously.~\cite{Bal2020}

It is worth pointing out that an earlier model of the droplet nucleation rate was also based on TST.~\cite{Schenter1999}
Compared to our approach, the TST rate expression was constructed only for a single (elementary) monomer addition or evaporation process at a cluster of fixed size $n$, whereas we derive a single expression for the total nucleation rate without explicitly assuming a mechanism based on sequential monomer addition only.

The FES is calculated within a small simulation cell, and the resultant barrier derived from this FES is the formation free energy of a critical nucleus (or, more generally, dividing surface) within this cell.
That is, the barrier in Eq.~\eqref{eq:barF} and TST rate Eq.~\eqref{eq:Eyring} are only defined for the $N$-atom system in which the nucleation free energy surface was obtained.
Therefore, $k^\mathrm{TST}$ is the TST nucleation rate inside this particular simulation cell.
To obtain a global nucleation rate $J$, we must divide $k$ by the initial volume $V$ of the simulation cell:
\begin{equation}
  J^\mathrm{TST} = \frac{k^\mathrm{TST}}{V} . \label{eq:JTST}
\end{equation}

$k^\mathrm{TST}$ is a strict upper bound to the true rate and, consequently, $J^\mathrm{TST}$ could overestimate the true nucleation rate $J$.
A failure of TST can mostly be traced back to one of two following phenomena:
\begin{enumerate}
  \item the reaction coordinate $n$ does not parametrize a proper dividing surface and $\Delta^\ddagger F$ is underestimated, or;
  \item not every crossing of the dividing surface $n = n^*$ results in an effective transition $g \rightarrow l$.
\end{enumerate}

\subsection{Committor analysis and recrossing correction}

We propose that committor analysis, which is a standard way to verify the quality of a candidate reaction coordinate $\chi$, can simultaneously be used to obtain a transmission coefficient $\kappa$, which compensates for recrossings of the dividing surface.
The committor $p_l$ is the probability that an ensemble of configurations commits to the liquid state $l$.~\cite{Geissler1999}
If $p_l > 0.5$ configurations can be considered to belong to the liquid state, if $p_l < 0.5$ they are part of the vapor, and if $p_l = 0.5$ they are part of the transition state ensemble.
As a result, the quality of our putative dividing surface $n = n^*$ as identified from the geometric FES \eqref{eq:GFES} can be assessed by subjecting a sample of $n = n^*$ states to a committor test.
Finding $p_l = 0.5$ is a necessary condition for being a suitable dividing surface and can thus be used to validate the reaction coordinate $n$.

If we find $p_l \approx 0.5$, we assume that recrossings are intrinsic to the true dividing surface.
We now also assume that the system spends such a long time in the transition state (TS) region that it becomes fully decorrelated.
This corresponds to fully diffusional barrier crossing dynamics.
The TST rate, by definition, amounts to all crossings of $n = n^*$.
Therefore, if we count the average number of TS crossings $j_\mathrm{cross}$ during the committor analysis we can directly measure the correction to the TST rate.
The fraction of TS crossings that effectively results in a nucleation event is $p_l / \langle j_\mathrm{cross} \rangle$.
This quantity therefore corresponds to the transmission coefficient $\kappa$.
In addition, if it was previously found that $p_l = 0.5$, we have $\kappa = (2 \langle j_\mathrm{cross} \rangle)^{-1}$.

The final estimate of the nucleation rate inside the cell volume $V$ is now
\begin{equation}
  k = \kappa k^\mathrm{TST} = \frac{\kappa}{h \beta} e^{-\beta \Delta^\ddagger F} ,
\end{equation}
and the global nucleation rate is
\begin{equation}
  J = \kappa J^\mathrm{TST} = \frac{\kappa k^\mathrm{TST}}{V} . \label{eq:J}
\end{equation}

Transmission coefficients and recrossings have received much interest, and several theories have been developed to rationalize the concept, including CNT or Kramers' theory.\cite{Kramers1940}
Here, we however only calculate numerical values of $\kappa$ for the chosen reaction coordinate, directly employing its definition within TST: The ratio between the effective rate and the TST rate associated with the reaction coordinate.
Compared to more dedicated theories, the current approach offers little direct physical insight but, as we will show, it is accurate and simple to apply.

\section{Computational details}

All simulations were carried out with LAMMPS~\cite{Plimpton1995} and the PLUMED plugin.~\cite{Tribello2014,PLUMED2019}
The interatomic interactions between the Ar atoms was described using a Lennard-Jones potential with $\epsilon = 0.99797$~kJ/mol and $\sigma = 3.405$~nm.
The interaction was truncated at a distance of $6.75 \sigma$.
These parameters fully match those used earlier.~\cite{Chkonia2009,Salvalaglio2016,Tsai2019}

The equations of motion were integrated with a time step of 5~fs and temperature control at $T = 80.7$~K was achieved using a Langevin thermostat~\cite{Bussi2007} with a time scale of 1~ps.
A Langevin thermostat was found to be necessary to maintain a strict equipartition of the energy in the system, between vapor and liquid phases.
Note that such a thermostat should not be used when explicitly sampling nucleation times (i.e., in brute force MD or infrequent metadynamics), since the Langevin friction affects the rate of processes.
For this reason, we used a global thermostat~\cite{Bussi2007Global} when performing committor analysis, which retains the equilibration efficiency of its local Langevin counterpart while leaving dynamical trajectories mostly unperturbed.~\cite{Bussi2008}

Constant volume (NVT) simulations were performed in periodic cubic simulation cells of different size.
Following previous definitions ~\cite{Chkonia2009,Salvalaglio2016,Tsai2019}, each system is identified by its supersaturation level $S$, defined as 
\begin{equation}
  S = \frac{N k_B T}{p_e V} , \label{eq:supersat}
\end{equation}
in which $p_e = 0.43$~bar.

Constant pressure (NPT) simulations were performed in the same way as the NVT simulations, except that the equations of motion are those of a Nos\'e--Hoover style barostat.~\cite{Martyna1994}
The imposed pressure $p$ for each value of $S$ is chosen by an initial NVT simulation in a box with a size found from eq.~\eqref{eq:supersat}.
We use this approach, rather than using $p = S p_e$, because we wish to comply with the ideal gas law-based naming scheme established earlier.

In order to achieve sufficient sampling along the reaction coordinate $n$, some enhanced sampling scheme is necessary.
Here, we choose well-tempered metadynamics~\cite{Laio2002,Barducci2008} because it is a widely available method that demonstrates the ease by which our approach can be implemented practically.
The approach presented here is however method-agnostic and other sampling strategies can employed to reconstruct the FES if they are more efficient or practical.
Other free energy studies of phase transitions have used, for example, umbrella sampling,~\cite{tenWolde1998,Auer2001,Auer2004,Cheng2018} variationally enhanced sampling (VES),~\cite{Piaggi2017,Piaggi2020} adiabatic free-energy dynamics (AFED),~\cite{Samanta2014,Rogal2019} on-the-fly probability-enhanced sampling (OPES),~\cite{Karmakar2021} and reweighted Jarzynski sampling.~\cite{Bal2021}

The metadynamics parameters were almost equal in all systems.
As a CV, we used the number of liquid atoms $n$, defined using switching functions of the type
\begin{equation}
  s(x) = \frac{1 - (x/x_0)^6}{1 - (x/x_0)^{12}} . \label{eq:switch}
\end{equation}
First, a coordination number $c_i$ is calculated for each atom $i$, by summing $s(r_{ij})$ using the pairwise distance $r_{ij}$ with all other atoms $j$ within 10~\AA{}, and $r_0 = 5$~\AA{}.
Then, $n$ is calculated as the sum of all $s(c_i)$, using $c_0 = 5$.

The external bias potential in metadynamics is history-dependent and expressed as a sum of repulsive Gaussians.
Every 50~ps, a Gaussian of initial height $w = 0.5$~kJ/mol was added to the total bias.
The width of each new Gaussian was determined using a diffusional scheme,~\cite{Branduardi2012} on a time scale of 25~ps.
Well-tempered metadynamics was used with bias factor $\gamma = 15$, to gradually reduce the size of the newly added Gaussians and smoothly converge the bias.~\cite{Barducci2008}
FES estimates were produced by reweighting~\cite{Tiwary2015} 200~ns chunks of the biased trajectory.
The total simulation time was 1~$\mu$s for each system.

We used harmonic restraints on $n$ to keep the droplet from growing too large.
These were placed at $n = 64$ (for $S > 6.01$) or $n = 128$ (otherwise).

Representative system configurations with $n = n^*$ for committor analysis were generated using steered MD (SMD), and a set of 10 independent trajectories were launched for each condition.
For each trajectory, we recorded the number of times $j_\mathrm{cross}$ the system crosses the dividing surface defined by $n = n^*$.
20~ns per trajectory proved to be sufficient for all systems, except for $S = 4.81$.

\section{Results and discussion}

\subsection{The finite size limit}

\begin{figure}[tb]
\includegraphics{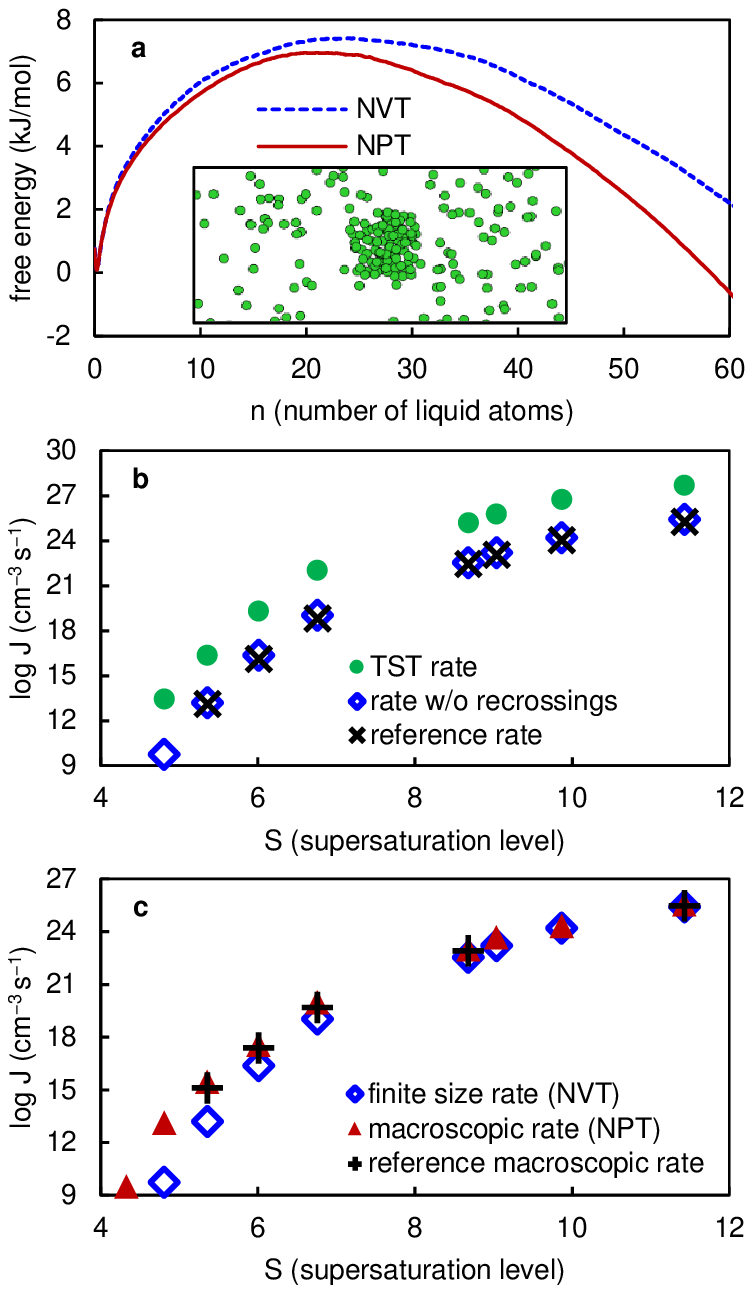}
\caption{\label{fig:panel-rates} Calculation of Ar droplet nucleation rates from TST.
(a) Effect of the ensemble on the free energy surface. Inset: snapshot of a droplet $n \approx 64$.
(b) Comparing computed TST rates $J^\mathrm{TST}$ with recrossing-corrected rates $J = \kappa J^\mathrm{TST}$ and literature results $J^\mathrm{ref}$ as a function of the supersaturation level $S$.
(c) Comparing rates $J$ in the finite system and NVT ensemble to their NPT counterparts $J_\infty$ and finite size-corrected macroscopic rates from the literature $J_\infty^\mathrm{ref}$ as a function of $S$.}
\end{figure}

We first  study nucleation in a vapor of 512 Ar atoms in the canonical (NVT) ensemble using Langevin dynamics~\cite{Bussi2007} within several fixed box volumes $V$ (Table~\ref{tab:nvt}).
Each box size represents a different supersaturation level $S$.
Specifically, we consider a series of systems at a temperature $T = 80.7$~K, for which accurate rate estimates are available from (accelerated) MD trajectories.~\cite{Chkonia2009,Salvalaglio2016,Tsai2019}
As an example, we plot in Fig.~\ref{fig:panel-rates}a the FES for nucleation in a cubic cell with an edge length of 11.5~nm, or $S = 8.68$.

\begin{table*}
\caption{\label{tab:nvt} Nucleation barriers $\Delta^\ddagger F$, TST rates $J^\mathrm{TST}$, transmission coefficients $\kappa$ and final rate estimates $J$ in a fixed-volume (NVT) 512 Ar system.
Different supersaturations $S$ are simulated by choosing the simulation cell edge length $L$.
Reference rates $J^\mathrm{ref}$ are given for comparison.\footnote{Reference values taken from Tsai et al.~\cite{Tsai2019} for $S \geq 9.04$ and from Salvalaglio et al.~\cite{Salvalaglio2016} otherwise.} }
\begin{ruledtabular}
\begin{tabular}{lcccccc}
S       & $L$   & $\Delta^\ddagger F$    & $J^\mathrm{TST}$               & $\kappa$        & $J$                            & $J^\mathrm{ref}$\\
        & (nm)  & (kJ/mol)               & (cm$^{-3}$ s$^{-1}$)           & ($10^{-3}$)     & (cm$^{-3}$ s$^{-1}$)           & (cm$^{-3}$ s$^{-1}$) \\       
\hline
11.43   & 10.5  & $3.76 \pm 0.15$        & $5.34 \pm 1.23 \times 10^{27}$ & $4.9 \pm 1.4$ & $2.62 \pm 0.95 \times 10^{25}$ & $1.84 \pm 0.27 \times 10^{25}$ \\
9.87    & 11.0  & $5.13 \pm 0.27$        & $6.03 \pm 2.44 \times 10^{26}$ & $2.6 \pm 0.9$ & $1.58 \pm 0.83 \times 10^{24}$ & $1.09 \pm 0.19 \times 10^{24}$ \\
9.04    & 11.3  & $6.59 \pm 0.18$        & $6.36 \pm 1.69 \times 10^{25}$ & $2.5 \pm 0.4$ & $1.61 \pm 0.50 \times 10^{23}$ & $1.10 \pm 0.27 \times 10^{23}$ \\
8.68    & 11.5  & $7.46 \pm 0.30$        & $1.65 \pm 0.73 \times 10^{25}$ & $2.1 \pm 0.5$ & $3.52 \pm 1.73 \times 10^{22}$ & $2.80 \pm 0.82 \times 10^{22}$ \\
6.76    & 12.5  & $12.20 \pm 0.19$       & $1.10 \pm 0.32 \times 10^{22}$ & $1.0 \pm 0.2$ & $1.10 \pm 0.39 \times 10^{19}$ & $0.64 \pm 0.33 \times 10^{19}$ \\
6.01    & 13.0  & $16.31 \pm 0.51$       & $2.12 \pm 1.63 \times 10^{19}$ & $1.1 \pm 0.2$ & $2.37 \pm 1.85 \times 10^{16}$ & $1.26 \pm 0.56 \times 10^{16}$ \\
5.36    & 13.5  & $20.78 \pm 0.13$       & $2.44 \pm 0.47 \times 10^{16}$ & $0.6 \pm 0.2$ & $1.57 \pm 0.48 \times 10^{13}$ & $1.30 \pm 0.75 \times 10^{13}$ \\
4.81    & 14.0  & $25.24 \pm 0.07$       & $2.81 \pm 0.29 \times 10^{13}$ & $0.2 \pm 0.1$ & $5.46 \pm 1.08 \times 10^9$    & \\
\end{tabular}
\end{ruledtabular}
\end{table*}

If we assume $\kappa = 1$, we can use \eqref{eq:Eyring} to calculate the TST rate $k^\mathrm{TST}$ and also obtain a TST-style estimate estimate $J^\mathrm{TST}$ of the global nucleation rate $J$.
Only the free energy surface $F(n)$ (and an appropriate gauge correction) is needed to calculate $k^\mathrm{TST}$.

While TS crossing in many chemical reactions can be considered to be ballistic (and $k \approx k^\mathrm{TST}$), this may not be the case for nucleation processes.
Not every occurrence of a configuration $\mathbf{R}$ for which $n(\mathbf{R}) = n^*$ necessarily corresponds to a nucleation event.
As can be seen in Fig.~\ref{fig:panel-rates}b and Table~\ref{tab:nvt}, a poor agreement with literature rates is obtained when we calculate a nucleation rate $J^\mathrm{TST}$ from $k^\mathrm{TST}$.
On average $J^\mathrm{TST}$ and $J^\mathrm{ref}$ deviate by three orders of magnitude.

From a committor test we always found that $p_l \approx 0.5$ in all cases, confirming the quality of the CV $n$.
From just 10 trajectories per $S$, we also obtained estimates of $\kappa$ that have a precision similar to that of $J^\mathrm{TST}$, and are in the order of $10^{-3}$ (Table~\ref{tab:nvt}).
TS crossing is therefore highly diffusive, thus validating our assumption that trajectories around $n = n^*$ become fully decorrelated.
Our final nucleation rate estimates $J$ now match very well the $J^\mathrm{ref}$ values, as can be seen in Fig.~\ref{fig:panel-rates}b.
This agreement is even more remarkable when realizing that rate estimates purely from CNT can be off by several orders of magnitude.~\cite{Diemand2013}
Such inconsistencies in nucleation rate predictions are quite common: A spectacular example is ice formation, for which rates calculated by different approaches (seeding, forward flux sampling, and a CNT-based recipe) were found to span nine orders of magnitude even though the employed water model and simulation conditions were the same.~\cite{Cheng2018}

\begin{figure}[t]
\includegraphics{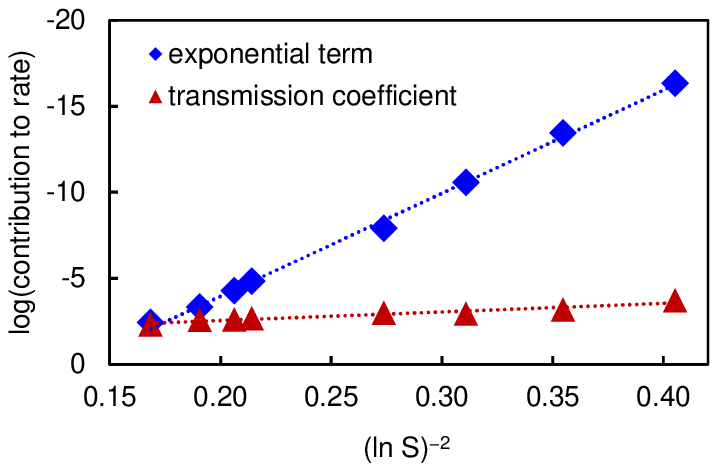}
\caption{\label{fig:panel-contribs} Relative contributions to the nucleation rate $J$.
Values of the exponential term $e^{-\beta \Delta^\ddagger F}$ and transmission coefficient $\kappa$ are plotted as a function of $\ln^{-2} S$.}
\end{figure}

As can be seen in Fig.~\ref{fig:panel-contribs}, the relative contribution of $\kappa$ to the overall nucleation rate is similar to that of the exponential $e^{-\beta \Delta^\ddagger F}$ term at high supersaturations.
With decreasing $S$, the nucleation barrier increases strongly, while $\kappa$ only has a weak dependence on $S$.
The fact that nucleation time scales over the whole studied supersaturation range span sixteen orders of magnitude can therefore be almost exclusively attributed to the exponential term in the rate expression Eq.~\eqref{eq:J}.

\subsection{Macroscopic nucleation rates}

Nucleation rates calculated in a small simulation box with fixed dimensions are affected by finite size effects.
This is because the growing droplet depletes the gas phase and, thus, artificially decreases the supersaturation.
Salvalaglio et al.~\cite{Salvalaglio2016} corrected their accelerated MD simulations by estimating the finite size error from CNT expressions.
However, a much more generic solution to the finite size problem is available within our FES-based approach.
If we calculate the FES within the constant pressure NPT ensemble, the vapor phase is kept at its initial pressure throughout the simulation because the box size is allowed to vary.

\begin{table*}
\caption{\label{tab:npt} Nucleation barriers $\Delta^\ddagger G$, TST rates $J^\mathrm{TST}_\infty$, transmission coefficients $\kappa$ and final rate estimates $J_\infty$ in a constant pressure (NPT) 512 Ar system that approximate the physics of a macroscopically sized system.
Different supersaturations $S$ are simulated by enforcing a pressure $p$.
Reference rates $J_\infty^\mathrm{ref}$ are given for comparison.\footnote{Reference values taken from Salvalaglio et al.~\cite{Salvalaglio2016} where available.} }
\begin{ruledtabular}
\begin{tabular}{lcccccc}
S       & $p$   & $\Delta^\ddagger G$    & $J_\infty^\mathrm{TST}$               & $\kappa$        & $J_\infty$                            & $J_\infty^\mathrm{ref}$\\
        & (atm) & (kJ/mol)               & (cm$^{-3}$ s$^{-1}$)           & ($10^{-3}$)     & (cm$^{-3}$ s$^{-1}$)           & (cm$^{-3}$ s$^{-1}$) \\       
\hline
11.43   & 3.92  & $3.65 \pm 0.17$        & $6.33 \pm 1.58 \times 10^{27}$ & $5.9 \pm 2.0$   & $3.75 \pm 1.56 \times 10^{25}$ & $3.04 \pm 0.70 \times 10^{25}$ \\
9.87    & 3.52  & $4.99 \pm 0.12$        & $7.44 \pm 1.32 \times 10^{26}$ & $2.9 \pm 1.0$   & $2.16 \pm 0.84 \times 10^{24}$ \\
9.04    & 3.30  & $6.12 \pm 0.07$        & $1.28 \pm 0.13 \times 10^{26}$ & $3.9 \pm 0.9$   & $5.05 \pm 1.25 \times 10^{23}$ \\
8.68    & 3.16  & $6.83 \pm 0.20$        & $4.22 \pm 1.24 \times 10^{25}$ & $2.4 \pm 0.4$   & $1.01 \pm 0.35 \times 10^{23}$ & $8.64 \pm 2.53 \times 10^{22}$ \\
6.76    & 2.55  & $11.47 \pm 0.38$       & $3.25 \pm 1.84 \times 10^{22}$ & $3.1 \pm 0.9$   & $1.00 \pm 0.64 \times 10^{20}$ & $0.51 \pm 0.27 \times 10^{20}$ \\
6.01    & 2.31  & $14.49 \pm 0.24$       & $3.20 \pm 1.13 \times 10^{20}$ & $1.2 \pm 0.4$   & $3.87 \pm 1.94 \times 10^{17}$ & $2.57 \pm 1.14 \times 10^{17}$ \\
5.36    & 2.08  & $17.42 \pm 0.06$       & $3.62 \pm 0.31 \times 10^{18}$ & $0.8 \pm 0.3$   & $2.92 \pm 1.04 \times 10^{15}$ & $1.35 \pm 0.78 \times 10^{15}$ \\
4.81    & 1.89  & $20.98 \pm 0.29$       & $1.62 \pm 0.69 \times 10^{16}$ & $0.9 \pm 0.2$   & $1.39 \pm 0.67 \times 10^{13}$ \\
4.33    & 1.71  & $26.23 \pm 0.13$       & $5.77 \pm 1.14 \times 10^{12}$ & $0.6 \pm 0.2$   & $3.32 \pm 1.13 \times 10^{9}$ \\
\end{tabular}
\end{ruledtabular}
\end{table*}

We therefore repeat our metadynamics simulations in the NPT ensemble.
Taking yet again the case of $S = 8.68$ as an example, we see that the FES of nucleation---which now represents the Gibbs free energy $G$ rather than the Helmholtz definition $F$---is significantly affected by the ensemble change (Fig.~\ref{fig:panel-rates}a and Table~\ref{tab:npt}).
In this system, the nucleation barrier $\Delta^\ddagger G_{g \rightarrow l}$ decreases by about 0.6~kJ/mol (${\sim} k_B T$).
$\kappa$ appears not appreciably affected by finite size effects, meaning that our final estimate of the macroscopic nucleation rate $J_\infty$ is about 3 times higher than the finite size estimate $J$, which is in perfect agreement with the estimates of Salvalaglio et al. (Fig.~\ref{fig:panel-rates}c).
More generally, our results closely match finite size-corrected nucleation rates for all $S$ with available reference data.
With decreasing $S$, the magnitude of the finite size effect increases very strongly, up till four orders of magnitude for $S = 4.81$.

It may also be possible to directly sample nucleation rates in the NPT ensemble, using accelerated MD.
However, although the employed thermo- and barostat correctly reproduce the thermodynamic averages of the target ensemble, they achieve this by augmenting the equations of motion with an artificial friction term.~\cite{Bussi2007,Martyna1994}
The dynamical trajectories of all atoms are thus affected.
It has therefore been argued that also nucleation times could be unphysical to some extent, although the magnitude of this possible effect was not quantified.~\cite{Diemand2013,Diemand2014}
In contrast, the TST rate is purely an equilibrium property of the system:
The FES (or barrier) only depends on the underlying thermodynamic distributions, and not the precise dynamical trajectories.

More generally, calculating nucleation free energy barriers is a matter of sampling along a suitable reaction coordinate (or CV), while maintaining the nucleating particles at a physically meaningful chemical potential $\mu$.
Depending on the process, such can be achieved in the NPT,~\cite{Quigley2009,Piaggi2017,Piaggi2020,Rogal2019,Pipolo2017,Zhang2019,Samanta2014,Gimondi2017} NVT,~\cite{Fukuhara2021} or $\mu$VT ensembles.~\cite{Karmakar2019}
As we show here, the resultant FES (and accompanying committor analysis) then suffices to calculate accurate macroscopic nucleation rates from TST.
Care must be taken, however, to ensure that the system is large enough to accommodate the critical nucleus.
Convergence tests using different system sizes can reveal any remaining size effects~\cite{Quigley2009} which, as will be shown in Section~\ref{sec:CNT}, are absent in our setup.

\subsection{Efficiency of the rate calculation}

The efficiency of accelerated MD simulations is often expressed in terms of an \emph{acceleration factor} $\alpha$, which is the ratio of the transition time $\tau$ and the length of the MD trajectory needed to observe it in the biased simulation, i.e., 
\begin{equation}
  \alpha = \frac{\tau}{t_\mathrm{MD}} . \label{eq:acc}
\end{equation}

The reference rates $J^\mathrm{ref}$ used here were obtained within the infrequent metadynamics framework.~\cite{Tiwary2013}
In infrequent metadynamics, a standard metadynamics setup is employed, but the deposition of the Gaussian bias potentials is done more slowly.
The idea is that this helps to ensure that the dividing surface between states (here $n = n^*$) is not biased, and the simulation becomes equivalent to hyperdynamics.~\cite{Voter1997}
Then, $\alpha = \langle e^{\beta V (n)}\rangle_b$, in which $V(n)$ is the bias potential and $\langle \cdots \rangle_b$ denotes a time average over the biased trajectory.

Using infrequent metadynamics simulations, $\alpha_\mathrm{iMetaD} = 1.7 \times 10^{11}$ could be reached for $S = 5.36$.
However, in order to obtain correct statistics, several independent   observations $n_\mathrm{sim}$ of the transition were needed.
We can use a similar definition to calculate $\alpha_\mathrm{FES}$ for the FES-based estimation of the rate, in which we take $t_\mathrm{MD}$ as the time needed to converge the FES estimates (1~$\mu$s in total), plus the time spent for committor analysis ($10 \times 20$~ns).
The discrepancy between the two definitions lies in the value of $n_\mathrm{sim}$.
Therefore, we can compute the relative efficiency $\eta$ of the two approaches as:
\begin{equation}
  \eta = n_\mathrm{sim} \frac{\alpha_\mathrm{FES}}{\alpha_\mathrm{iMetaD}} . \label{eq:relacc}
\end{equation}

\begin{table}
\caption{\label{tab:eff} Nucleation times $\tau$ in NVT simulations, number of infrequent metadynamics runs $n_\mathrm{sim}$ and acceleration factors $\alpha_\mathrm{iMetaD}$ in the literature,\footnote{$\alpha_\mathrm{iMetaD}$ and $n_\mathrm{sim}$ values taken from Tsai et al.~\cite{Tsai2019} for $S \geq 9.04$ and from Salvalaglio et al.~\cite{Salvalaglio2016} otherwise.} acceleration by the FES-based approach $\alpha_\mathrm{FES}$, and relative efficiency $\eta$ of the two. }
\begin{ruledtabular}
\begin{tabular}{lccccc}
S       & $\tau$ (s)             & $n_\mathrm{sim}$  & $\alpha_\mathrm{iMetaD}$       & $\alpha_\mathrm{FES}$        & $\eta$    \\
\hline
11.43   & $3.30 \times 10^{-8}$  & 20                & $6.40 \times 10^{1}$           & $2.75 \times 10^{-2}$        & 0.01 \\
9.87    & $4.77 \times 10^{-7}$  & 20                & $1.10 \times 10^{3}$           & $3.97 \times 10^{-1}$        & 0.01 \\
9.04    & $4.29 \times 10^{-6}$  & 20                & $7.80 \times 10^{3}$           & $3.58 \times 10^{0}$         & 0.01 \\
8.68    & $1.87 \times 10^{-5}$  & 100               & $1.80 \times 10^{2}$           & $1.56 \times 10^{1}$         & 9.95 \\
6.76    & $4.64 \times 10^{-2}$  & 50                & $2.40 \times 10^{5}$           & $3.87 \times 10^{4}$         & 8.06 \\
6.01    & $1.92 \times 10^{1}$   & 50                & $6.30 \times 10^{7}$           & $1.60 \times 10^{7}$         & 14.44 \\
5.36    & $2.59 \times 10^{4}$   & 50                & $1.70 \times 10^{11}$          & $2.16 \times 10^{10}$        & 11.66 \\
4.81    & $6.67 \times 10^{7}$   &                   &                                & $5.56 \times 10^{13}$        & \\
\end{tabular}
\end{ruledtabular}
\end{table}

A clear discrepancy between different infrequent metadynamics studies becomes apparent, where Tsai et al. used significantly more aggressive biasing parameters than Salvalaglio et al.
In addition, and also not directly discernable from Table~\ref{tab:eff}, the former authors used a more complex CV, which besides $n$ also contained information about droplet shape.
A single MD step in the study of Tsai et al. therefore also required more CPU time.
A short test indicated that our implementation of $n$ is about 30 times faster to evaluate than their preferred CV for biasing.
Indeed, due to the high cost of their simulations no supersaturations below 9.04 could be simulated.~\cite{Tsai2019}

Nevertheless, we see that a FES-based approach only starts to become competitive at lower supersaturations, where it quite consistently outperforms infrequent metadynamics by an order of magnitude.
In addition, without changing biasing parameters, we could calculate rates for supersaturations as low as $S = 4.33$ in the NPT simulations, with $\tau \sim 10^8$~s, and $\alpha_\mathrm{FES} = 8.2 \times 10^{13}$. 
We can therefore anticipate that using the FES and TST becomes an increasingly attractive option when interatomic potentials become more expensive and/or nucleation barriers become higher.

Furthermore, metadynamics may not necessarily be the most efficient free energy method under all conditions.
Plenty of alternative free energy methods have been reported in the literature and implemented in widely available codes such as PLUMED.
We have recently already demonstrated that a new method based on nonequilibrium sampling improves upon metadynamics by a factor ${\sim}3$ in the $S = 8.68$ NVT case.~\cite{Bal2021} 

\subsection{Discussion of errors}

Overall, the agreement between infrequent metadynamics and the TST-based approach is very good.
This is quite remarkable considering that the most prominent sources of error of both methods go in opposite directions.

Suboptimal CVs have a negative impact on the performance of infrequent metadynamics: If the CV does not contain all slow modes in the system, the bias potential will not be effective, leading to overfilling of the metastable basin before a transition can occur.
Or, put differently, a poor CV will not properly distinguish transition states from metastable states, meaning that bias is also added to the transition states, leading to a violation of the hyperdynamics assumption.~\cite{Voter1997}
As a result, transition times will be overestimated, and predicted rates will be \emph{underestimated}.~\cite{Tsai2019,Khan2020}

A poor CV can still be sufficient to enhance sampling and converge a FES.
Because it mixes TS states with stable states the apparent free energy of the TS will however be too low.
Therefore, rates computed from this barrier will always be an upper bound of the true rate, and are prone to \emph{overestimate} it.

We attempted to minimize the error in the reference nucleation rates by selecting the values Tsai et al. obtained using their optimized CV, rather than $n$.
Salvalaglio et al. only used $n$, but were significantly more prudent with respect to their biasing parameters.

Despite using $n$ as a CV, which may be suboptimal~\cite{Tsai2019}, our rates are very close to the infrequent metadynamics estimates.
There may, however, be a slight bias to somewhat higher rates (up to 2 times higher than the reference, but always within error bars), in line with the reasoning outlined above.
The overall good agreement of the competing approaches is however consistent with the observation of Tsai et al. that the barrier along their optimized CV was not appreciably higher than the one along $n$.~\cite{Tsai2019}
Note, also, that small differences in numerical precision between the employed codes may introduce small deviations.

Finally, our committor analysis reveals one important point of caution when applying infrequent metadynamics along $n$.
Because the system may spend up to 10~ns in the TS region, it is very difficult to guarantee an uncorrupted TS if new Gaussian biases are continuously added: Only simulations with impractically low bias addition rates or very small Gaussians are truly trustworthy.

Rates have an exponential dependence on nucleation free energy barriers, so even small uncertainties in the FES can result in large error bars on a final rate estimate.
For almost every system we have managed to keep the uncertainty on the barrier well below $k_B T$, leading mostly to errors between 30 and 60~\% on the rate.
These error bars are similar to those reported by Salvalaglio et al.~\cite{Salvalaglio2016}
Somewhat higher uncertainties have been reported on nucleation rates computed by forward flux sampling (between 150 and 500~\%) in diverse systems.~\cite{Wang2009,Haji-Akbari2015,Sosso2016}

\subsection{Interpretation of the transmission coefficient}

In the strictest sense, the objective of our study is to obtain nucleation rates.
As a consequence we have not interpreted the values of the nucleation barrier or transmission coefficient $\kappa$ in great detail.
These two quantities are however quite interconnected.

Indeed, the value of $\kappa$ as used in our study can potentially serve two purposes.
If we assume that $n = n^*$ is the best possible choice of dividing surface, the free energy barrier will be maximized (in the spirit of variational TST) and $\kappa < 1$ represents the inherent diffusivity in the TS region.
In such case, a no-recrossing dividing surface does not exist, and $\kappa$ captures dynamical (friction) effects that lower the true TST rate.

Alternatively, recrossings may also be a consequence of a poorly chosen dividing surface.
In that case $n = n^*$ also contains configurations with lower free energies (so the apparent barrier is too low) and is crossed more than the true dividing surface (so $\kappa$ becomes smaller).
In such a situation, the final rate estimate may still be accurate, but barrier and transmission coefficient have less of a clear-cut physical significance.

However, note that we use steered MD to generate trial $n = n^*$ configurations for committor analysis, starting in the $g$ state.
If $n = n^*$  is a poor dividing surface, one would expect these configurations to be biased towards the $g$ state because there is a residual barrier still unaccounted for in $F(n)$.
As a result, the SMD run will be unable to place the system exactly on the true dividing surface and $p_l < 0.5$ 
Because we do, however, find $p_l \approx 0.5$ we conclude that such residual barrier is negligible (${\sim} k_B T$).
More rigorous tests for candidate dividing surfaces have also been proposed.~\cite{Mullen2014}

It is therefore likely, then, that the very low values of $\kappa$ (in the order of $10^{-3}$) are mostly a manifestation of intrinsic dynamic effects.
This is a reasonable conclusion, considering that droplet growth is a process fully driven by diffusion of gas atoms, balanced by re-evaporation of atoms from the liquid.
The stochastic nature of these phenomena is compatible with small transmission coefficients.

\subsection{Comparison with a CNT-based approach}
\label{sec:CNT}

Our method bears some resemblance with the popular ``parameter-free'' implementation of CNT pioneered by Auer and Frenkel.~\cite{Auer2001,Auer2004}
As in our approach, a nucleation FES must be reconstructed first and a rate estimate is calculated from the barrier.
It is important to note, however, that this expression is based on a definition of the barrier within the macroscopic CNT framework.

To wit, within the NPT ensemble, the rate is expressed as
\begin{equation}
  J = Z f^+ \rho_g e^{-\beta G^*} . \label{eq:CNT}
\end{equation}
Herein, $G^* = G(n^*) - G(0)$, $\rho_g$ is the number density of the metastable vapor, $f^+$ is the attachment rate on the critical nucleus, and $Z$ is the Zeldovich factor.

The attachment rate $f^+$ can be calculated in several ways.
Most commonly, one launches several trajectories starting from a critical nucleus, and calculates $f^+$ as a diffusion coefficient in $n$:
\begin{equation}
  f^+ = \frac{\langle (n(t) - n(0))^2 \rangle}{2t} . \label{eq:fmicro}
\end{equation}
Alternatively, in the case of droplet nucleation, one can use kinetic gas theory~\cite{Salvalaglio2016}:
\begin{equation}
  f^+ = A(n^*) \frac{\rho_{g,e}}{\sqrt{2 \pi \beta m}} . \label{eq:fmacro}
\end{equation}
The density of the vapor at coexistence is $\rho_{g,e} = \rho_g / S$.
$A(n^*)$ is the surface area of the critical nucleus, which can be calculated if the number density $\rho_l$ of the liquid is known:
\begin{equation}
  A (n^*) = \left(\frac{36 \pi}{\rho_l^2}\right)^{1/3} (n^*)^{2/3} .
\end{equation}

The Zeldovich factor $Z$ is computed from the free energy surface $G (n)$ as
\begin{equation}
  Z = \sqrt{\frac{\beta}{2\pi} \left| \frac{\mathrm{d}^2 G (n)}{\mathrm{d} n^2} \right|_{n=n^*}} .
\end{equation}

We illustrate the application of this approach with a calculation of $J_\infty$ for $S = 8.68$.
From the FES $G(n)$ we compute $Z = 0.065$.
Our two possible estimates of $f^+$ however differ quite strongly: 
The direct measurement of the diffusion coefficient using Eq.~\eqref{eq:fmicro} yields $2.5 \times 10^{11}$~s$^{-1}$, whereas the kinetic gas theory expression Eq.~\eqref{eq:fmacro} predicts $9.2 \times 10^9$~s$^{-1}$.
Also note that the former estimate is difficult to converge, and has a relative error bar of 100\%.

\begin{figure}[t]
\includegraphics{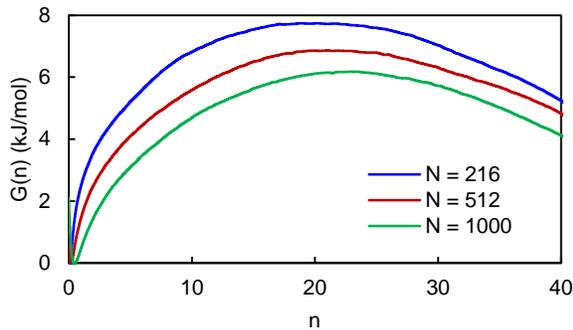}
\caption{\label{fig:panel-sizedep} Size dependency of the nucleation free energy surface $G (n)$ for $S = 8.68$.
Different system sizes $N$ are compared.}
\end{figure}

\begin{table}
\caption{\label{tab:bars} Size-dependence of nucleation barriers obtained from NPT simulations with different number of atoms $N$, for $S = 8.68$.\footnote{All barriers were computed from a single, long reweighting run. For fairness, this is also true for the $N=512$ case. This explains the slightly different value for $\Delta^\ddagger G$ compared to the value in Table~\ref{tab:npt}. It also explains the lack of error bars.}
Values of $J_\infty^\mathrm{TST}$, directly estimated from Eqs.~\eqref{eq:Eyring} and \eqref{eq:JTST}, show that our TST-based procedure naturally accounts for the extensive nature of $\Delta^\ddagger G$.}
\begin{ruledtabular}
\begin{tabular}{lccc}
$N$       & $G(n^*) - G(n_\mathrm{min})$     & $\Delta^\ddagger G$  & $J_\infty^\mathrm{TST}$ \\      
          & kJ/mol                           & kJ/mol               & cm$^{-3}$~s$^{-1}$ \\
\hline
216       & 7.74                             & 7.26                 & $5.2 \times 10^{25}$ \\
512       & 6.86                             & 6.68                 & $5.3 \times 10^{25}$ \\
1000      & 6.18                             & 6.36                 & $4.4 \times 10^{25}$ \\
\end{tabular}
\end{ruledtabular}
\end{table}

Now, we must calculate $G^* = G(n^*) - G(0)$.
This expression is however only valid if $G(n)$ has the shape predicted by CNT.
Because the order parameter $n$ does not strictly count the number of atoms in the critical nucleus only, $G(n)$ thus deviates from the CNT shape in particular for small $n$ with increasing system size $N$ (Fig.~\ref{fig:panel-sizedep}).
The minimum of this curve is now located at $n_\mathrm{min} > 0$
Therefore, the apparent barrier height $G(n^*) - G(n_\mathrm{min})$ is also size-dependent (Table~\ref{tab:bars}).
The precise nature of these issues was only recently addressed in full detail.~\cite{Yi2012,Cheng2017}
In principle, the definition of the barrier as $G(n^*) - G(0)$ can be retained only if $G(n)$ is transformed first into a macroscopic function consistent with the CNT definition of critical cluster size.~\cite{Cheng2017}

A more ad hoc correction can be derived as follows.
$e^{-\beta G^*}$ is defined as the relative ``equilibrium'' probability to form a critical nucleus around one monomer, and has to be multiplied by the monomer density $\rho_g$ to yield the ``equilibrium concentration'' of critical nuclei.
It is therefore an intensive, macroscopic quantity.
The quantity $G(n^*) - G(n_\mathrm{min})$ is the work required to form a critical nucleus inside the the simulation cell, i.e., the form a critical nucleus around \emph{any} of the monomer particles.
It is therefore an extensive quantity.
$e^{-\beta (G(n^*) - G(n_\mathrm{min}))}$ is therefore the probability of finding a critical nucleus within the simulation cell volume, relative to the system residing exactly in its local minimum.
The critical nucleus concentration therefore equals $e^{-\beta (G(n^*) - G(n_\mathrm{min}))}/V$.
If a nucleation barrier is obtained from microscopic simulations, one can therefore approximate the term $\rho_g e^{-\beta G^*}$ by $e^{-\beta (G(n^*) - G(n_\mathrm{min}))}/V$ in Eq.~\eqref{eq:CNT}, as noted before.~\cite{Yi2012}
Alternatively, $\beta G^* \approx \beta G(n^*) - \beta G(n_\mathrm{min}) + \ln N$.

When now using Eq.~\eqref{eq:fmicro} to estimate $f^+$, employing an appropriate definition of $G^*$, we find a predicted rate of $J_\infty \approx 4 \times 10^{23}$~cm$^{-3}$~s$^{-1}$, which is quite close to our TST-based result of $1.0 \times 10^{23}$~cm$^{-3}$~s$^{-1}$ when taking into account the very large uncertainty on $f^+$.
Eq.~\eqref{eq:fmacro} fares worse, yielding an estimated $J_\infty \approx 1.4 \times 10^{22}$~cm$^{-3}$~s$^{-1}$.
These results highlight that TST, CNT, and related approaches are equivalent theories that can be used to calculate nucleation rates.

Application of the CNT-derived expression Eq.~\eqref{eq:CNT} thus requires some processing to turn the microscopic simulation data into appropriately macroscopic quantities.~\cite{Cheng2017,Yi2012}
The TST rate of Eq.~\eqref{eq:kTST}, in contrast, is one monolithic expression for the flux through the dividing surface $n = n^*$.
It is therefore a purely microscopic quantity that is rigorously defined \emph{within the chosen simulation cell}.
This local rate estimate can then straightforwardly be converted in a global nucleation rate (through Eq.~\eqref{eq:JTST} or \eqref{eq:J}), which is also an experimentally verifiable observable.
In this sense the barrier $\Delta^\ddagger G$ and transmission coefficient $\kappa$ only have significance for the specific simulation setup in which they were obtained; they serve as input for our procedure to yield the final nucleation rate estimate $J$.

$\Delta^\ddagger G$ as defined by Eq.~\eqref{eq:barF}, in particular, does not correspond to the macroscopic nucleation barrier $G^*$ of Eq.~\eqref{eq:CNT} because it is also size-dependent (Table~\ref{tab:bars}).
Application of Eqs.~\eqref{eq:kTST} and \eqref{eq:JTST}, however, takes care of producing a macroscopic quantity.
It can also be seen that no lingering size effects remain in nucleation rate estimates from our procedure because the predicted TST nucleation rate $J_\infty^\mathrm{TST}$ is the same (within error) for each system size (Table~\ref{tab:bars}).

\section{Conclusions}
Enhanced sampling methods and TST provide a unified theoretical framework for rate calculations.
Whenever it is possible to converge a FES along a suitable approximate reaction coordinate, accurate rates can be computed at little extra cost.
Here, we have used Ar droplet nucleation as an example.
Global, macroscopic, nucleation rates can be unambiguously obtained from small model systems without the need to invoke a process-specific approximation such as CNT, as long as an appropriate ensemble is simulated.

Only two ingredients are required in a TST-based nucleation rate calculation.
Calculation of the TST rate requires the \emph{free energy barrier}, which can be obtained through an ever-increasing array of free energy methods.
The quality of the free energy barrier can subsequently be verified from a committor analysis of the candidate transition state, which yields an accurate \emph{recrossing correction} (i.e., transmission coefficient) as a byproduct.
Accurate, consistent, and reproducible nucleation rates are thus accessible through a straightforward application of widely available, well-tested and actively developed tools.

Although we have focused on nucleation process, the highly generic nature of the approach likely makes it conveniently straightforward to apply to any type of process in chemistry, materials science, and biology.

\begin{acknowledgments}
K.M.B. was funded as a junior postdoctoral fellow of the FWO (Research Foundation -- Flanders), Grant 12ZI420N.
The computational resources and services used in this work were provided by the HPC core facility CalcUA of the Universiteit Antwerpen, and VSC (Flemish Supercomputer Center), funded by the FWO and the Flemish Government.
K.M.B. thanks Erik Neyts for his continuous support.
\end{acknowledgments}

\section*{Author Declarations}
\subsection*{Conflict of interest}
The author has no conflicts to disclose.

\section*{Data availability}
The data that support the findings of this study are available from the corresponding author upon reasonable request.
Sample inputs to reproduce the reported simulations are deposited on PLUMED-NEST (www.plumed-nest.org), the public repository of the PLUMED consortium~\cite{PLUMED2019}, as plumID:21.009.\cite{data}

\bibliography{bibliography}

\end{document}